\documentclass[conference]{IEEEtran}
\IEEEoverridecommandlockouts
\usepackage{cite}
\usepackage{amsmath,amssymb,amsfonts}
\usepackage{algorithmic}
\usepackage{graphicx}
\usepackage{textcomp}
\usepackage{balance}
\usepackage{ltxtable}
\usepackage[table]{xcolor}  

\definecolor{mycolor}{rgb}{1,0.0,0.5}

\newtheorem{theorem}{Theorem}
\newtheorem{lemma}{Lemma}
\newtheorem{definition}{Definition}

\def\BibTeX{{\rm B\kern-.05em{\sc i\kern-.025em b}\kern-.08em
    T\kern-.1667em\lower.7ex\hbox{E}\kern-.125emX}}

\begin{document}

\title{Optimal Association Strategy of Multi-gateway Wireless Sensor Networks Against Smart Jammers 
}

\author{\IEEEauthorblockN{Mohammad Reza Heidarpour$^*$ and Mohammad Hossein Manshaei$^{\dag}$}
\IEEEauthorblockA{\textit{$^*$Department of Electrical and Computer Engineering, Isfahan University of Technology, Isfahan, Iran} \\
\textit{$^{\dag}$Department of Electrical and Computer Engineering, University of Arizona, Tucson, AZ, USA}}
}

\maketitle

\begin{abstract}
Engineers have numerous low-power wireless sensor devices in the current network setup for the Internet of Things, such as ZigBee, LoRaWAN, ANT, or Bluetooth. These low-power wireless sensors are the best candidates to transfer and collect data. But they are all vulnerable to the physical jamming attack since it is not costly for the attackers to run low power jammer sources in these networks. Having multiple gateways and providing alternative connections to sensors would help these networks to mitigate successful jamming. In this paper, we propose an analytical model to solve the problem of gateway selection and association based on a Stackelberg game, where the jammer is the follower. We first formulate the payoffs of both sensor network and attacker and then establish and prove the conditions leading to NASH equilibrium. With numerical investigation, we also present how our model can capture the performance of sensor networks under jamming with a varying number of gateways. Our results show that compared to the single gateway scenario, the network's throughput will improve by 26\% and 60\% when we deploy two and four gateways in the presence of a single jammer.    
\end{abstract}

\begin{IEEEkeywords}
 Internet of Things (IoT), wireless sensor networks, jamming attack, Stackelberg game
\end{IEEEkeywords}

\section{Introduction}
Having passed through all the rises  and downs, the wireless sensor network (WSN) technology is now in its plateau of productivity with enormous applications in the Internet of things (IoT) era \cite{kocakulak2017overview}. 
The WSNs are typically   consist of a number of low cost sensors with small, if any, infrastructure support. This ``light-weight"  feature is the winning bid of WSNs to be adopted in many use-cases such as tracking and monitoring, among others \cite{YICK20082292}. 
\\ One of the challenges of WSNs is to maintain the ``battery-powered" sensors  at the lowest possible expense while prolonging the network life time. Many advances in electronics such as energy-harvesting and RFID circuits have been emerged to respond to this demand. On the other hand, communication protocols, such as   BLE,  Zigbee, 6TiSCH, and LPWAN  (Sigfox, LoRa, NB-IoT, LTE-M etc.), have been introduced for radio energy  conservation. However, signaling at low power (\textit{whispering}), limited resources, and distributed nature make  WSNs also vulnerable to security threads, such as flooding, Denial-of-Service (DoS), jamming, and Sybil attacks \cite{butun2019security}. In particular, due to low power operation of sensors, jamming attacks to WSNs are inexpensive using commodity hardwares \cite{aras2017selective}.   
\\
In jamming attacks, the malicious device (\textit{jammer}) disturbs the communication link(s) by sending interfering signals. 
Jamming techniques can take different forms varying in the implementation complexity, energy efficiency, and stealthiness \cite{pelechrinis2010denial}. 
In \textit{active} jamming, the jammer sends its interfering signals, regardless of whether any legitimate user is transmitting on the channel or not.
On the other hand, in \textit{reactive} jamming, the jammer first senses the channel to detect/predict when legitimate signals are being
transmitted, and then, it may jam the channel.\\ 
The authors in \cite{namvar2016jamming, 9380308} studied an OFDM-based downlink scenario in which a jammer allocates its (interfering)
power over different subcarriers in a manner that the
 number of affected devices is maximized. On the other hand, as a countermeasure,
the IoT controller can also tune its power allocation across subcarriers to relieve the IoT devices as much as possible. As both the jammer and the controller play with the same resource, i.e., the power across subcarriers, the problem of finding the best strategy is modeled as a Colonel
Blotto game. 
In \cite{yang2013coping}, the jammer first learns the transmission power among a pair of WSN's users and  adjusts its own power for the maximum damage. The problem of finding the optimum power is approached by modeling the problem as a Stackelberg game, and finding its equilibrium.
A cognitive radio network under the attack of a smart jammer with  inaccurate  observation  on the power of  the secondary user (SU)  transmissions has been  studied in \cite{7076591}. Again, this is the transmit power that the jammer/SU may change for the attack/countermeasure and the problem is formulated  as a Stackelberg  game. 
In \cite{chiariotti2019game}, authors highlight the importance of an intelligent defense against jamming attacks by considering a jamming attack on the WSNs which targets both the disruption of the communication and depletion of the sensor's energy.  The attacker and the defender (sensor) are modeled as the players in an asymmetric Bayesian game. The optimal and heuristic strategies, for the attacker and the defender, are derived which determine how much energy to assign to each transmission.
\\
In general, large scale WSNs may have more than one gateway, following a hierarchical architecture: a tier of sensor nodes, and a tier of gateway nodes \cite{abbasi2007survey}. To the best of our knowledge, the problem of jamming attacks has not yet been considered in \emph{multi-gateway} WSNs. Aside from power control, the multi-gateway networks possess a new degree of freedom to combat the jamming attack through suitable (i.e., jamming-aware) sensors-to-gateways association. Thus, the focus of our work in this paper is on the problem of finding the optimal association strategy which
maximizes the average packet delivery rate (PDR) under jamming attacks.
The jammer can operate simultaneously in multiple channels, reactively and selectively. In particular, the jammer, in deciding to jam which sensors, is capable of sensing multiple uplink (sensor-to-gateway) channels , \textit{predicting} the outcome of each choice on the WSN's PDR, and select the most adversarial action (jamming those channels (sensors) with the worst effect). Our contribution is summarized as follows.
\begin{itemize}
	\item We propose a general (agnostic to PHY/MAC protocols) probabilistic model to abstract the operation of multi-gateway WSNs in jamming-poisoned environments. Our proposed model considers the locality of the jammers which act based on their measurements.  
	\item We formulate the problem of finding the best strategy for the sensor-to-gateway association under the smart jamming attack as finding the  equilibrium point of a \emph{Stackelberg game}.  
	\item We derive the equilibrium point of the proposed game, which is the solution of a bilevel integer non-linear programming (BINLP). We further reduce the BINLP to an integer linear programming (ILP) which can be fed to general-purpose ILP solvers (e.g., intlinprog, CPLEX, and Gurobi). 
	
\end{itemize}

We believe that our model can facilitate the analysis of long-term strategic interactions against jammers. This model would be the first step towards a deeper analysis of the learning approach, where we need to obtain more information about an unknown attacker. The rest of the paper is organized as follows. In Section~\ref{sec_SysModel}, the system model is presented and the problem is formulated as a Stackelberg game. In Section III, we derive an ILP to solve the game and find the WSN's best association strategy. Simulation results are provided in Section IV. Finally, Section V concludes this paper.

\section{System Model And Game Formulation}
\label{sec_SysModel}
In this section, first we present the system model, and the constraints and the objectives of the WSN and the jammer. We then formulate the WSN's association problem in the presence of a smart jammer as a Stackelberg game.

\subsection{System Model}
\label{sub_SysModel}

 \begin{figure}[t]
 \centering
 \includegraphics [scale=1.25]{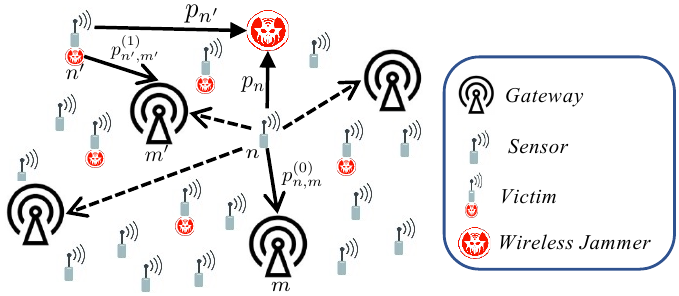}
\caption{Our system model includes a set of $N$ sensor nodes (SNs) and $M$ gateways (GNs). The main goal of the jammer is to reduce the average  uplink packet delivery rate of the network by jamming links between SNs and the associated GNs. In this example, SN $n$ has four feasible connections to four GNs and selects GN $m$. SN $n'$ is under jamming attack. $p_n$ is the probability that jammer detects the signal of SN $n$.  $p^{(0)}_{n,m}$ and $p^{(1)}_{n',m'}$ designate the probability of successful transmission for SN $n$ and SN $n'$ in this scenario. }
\label{fig:SysModel}
\end{figure}

Figure~\ref{fig:SysModel} presents our system model and Table~\ref{tab:TxModelSymbols} summarizes the notations used throughout the paper. We consider a WSN consisting of $M$ gateway nodes (GN) and $N$ sensor nodes (SN). The GNs are responsible to collect data from SNs, possibly process, and transmit the (extracted) information to a remote controlling entity. Each SN $ n\in\mathcal{N}=\{1,2,\cdots,N\} $ requires to be associated to one GN $ m\in\mathcal{M}=\{1,2,\cdots,M\} $, to which it directly sends its packets. The problem of finding the optimal SN-to-GN association, which maximizes the average packet delivery rate (PDR), can be formulated as: 
\begin{align}
\label{wsnp}
\begin{array}{llr}
\mathbf{P1:}& \max_{\mathbf{x}} \sum_{m=1}^{M}\sum_{n=1}^{N}p_{n,m}x_{n,m}&\vspace{1mm}\\\vspace{1mm}
&\text{subject to}\\
&\quad
\sum_{m=1}^{M}x_{n,m}\leq1,\quad\quad\hspace{0.5mm} n\in\mathcal{N}&\quad\text{(\ref{wsnp}-1)}\\
&\quad
\sum_{n=1}^{N}x_{n,m}\leq \chi_{m},\quad\quad\hspace{-1.5mm} m\in\mathcal{M}&\text{(\ref{wsnp}-2)}\\
&\quad
\sum_{m=1}^{M} c_my_m
\leq \Gamma,&\text{(\ref{wsnp}-3)}\\
&\quad
x_{n,m}\leq y_m,\quad\quad\hspace{1mm}\quad\quad m\in\mathcal{M}, ~n\in\mathcal{N}&\text{(\ref{wsnp}-4)}\\
&\quad
y_m,x_{n,m}\in\{0,1\}\quad\quad m\in\mathcal{M}, ~n\in\mathcal{N}&\text{(\ref{wsnp}-5)}\\
\end{array}
\end{align} 
where $\mathbf{x}=[x_{1,1},x_{1,2}, and \cdots,x_{N,M}]$,  $x_{n,m}\in\{0,1\}$ is a binary variable indicating whether the SN $ n $ is associated to GN $ m $ ($x_{n,m}=1$) or not ($x_{n,m}=0$).
 $p_{n,m}$ is the probability of successful packet transmission from the SN $ n $ to GN $ m $, $ \chi_{m} $ is the maximum number of SNs that can associate to GN $ m $. The problem  \textbf{P1} also considers  operational expenditure by including a cost of $c_m$ for turning GN $m$ on and the parameter $\Gamma$ as the maximum average operational cost for all gateways. Moreover, $y_m\in\{0,1\}$ shows the on-off sate of the GN $ m $. By tuning $\Gamma$, the number of GNs which are on can be controlled.

 \begin{table}[t]
 	\caption{List of Symbols in our Model}
 	\label{tab:TxModelSymbols}
 	\centering
 		\rowcolors{2}{gray!25}{white}
 		\begin{tabular}{|p{1.3cm}|p{6.3cm}|}
 				\rowcolor{gray!50}
 			 \hline 
 			 \hfil \textbf{Symbol} & \textbf{Definition}\\
 			   \hfil   $N$ &  Number of nodes \\
 			   \hfil   $M$ &  Number of Gateways \\
 			   \hfil   $x_{n,m}$ &  Association variable ($1$ if SN $n$ is connected to $m$) \\
 			   \hfil   $y_{m}$ &  State of GN ($1$ if GN $m$ is on) \\
 			   \hfil $ \chi_{m} $ & The maximum number of associated SNs to GN $m$ \\
 			   \hfil $\Gamma$ & The maximum average operational cost for all gateways \\
 			   \hfil   $p^{(0)}_{n,m}$ &  Probability of Successful Packet Transmission with no Jamming Attack (from SN $n$ to GN $m$)\\
 			   \hfil   $p^{(1)}_{n,m}$ & Probability of Successful Packet Transmission Under Jamming Attack (from SN $n$ to GN $m$) \\
			   \hfil  $p_n$ &  Probability of Detecting the Signal (sent by SN $n$) by the Jammer \\
			   \hfil $p_{n,m}$ & Probability of Successful Packet Transmission (from SN $n$ to GN $m$)\\
			   \hfil   $\tilde{p}^{(0)}_{n,m}$ &  $p^{(0)}_{n,m}$ From the Jammer's Perspective\\
			   \hfil   $\tilde{p}^{(1)}_{n,m}$ &  $p^{(1)}_{n,m}$ From the Jammer's Perspective\\
 			   \hfil   $v_n$ & Victim State of SN ($1$ if SN $n$ is selected as a victim by the jammer) \\	
 			   \hfil   $w_n$ & Sensor Importance Parameter from the Jammer's Perspective (the higher the absolute value, the more important the SN $n$ is)\\
 			   \hfil $\Omega$ & Power of the Jamming Signal\\
 			   \hfil   $\rho_n$ &  Average Consumed Power to Jam SN $n$ (if it is selected as a victim)\\
 			   \hfil   $\lambda$ &  Weight Balancing the Adversary Effect and the Amount of Power Usage in the Jammer's Objective Function\\
 			    \hfil   $a_{n,m}$ &  Association Penalty Parameter (w.r.t. SN $n$ to GN $m$) \\
 			     \hfil $z_{n,m}$& Victim-Association Variable (1 if SN $n$ is a victim and associated to GN $m$)\\
 			     \hfil   SPKT$( n )$ & Successful Packet Transmission (sent by SN $n$) \\
 			     \hfil   SPKT$( n,m )$ & Successful Packet Transmission (sent by SN $n$ to GN $m$) \\
 			   \hline
 		\end{tabular}
\end{table}
In \textbf{P1}, constraint set (\ref{wsnp}-1) ensures that  each SN is at most associated to one GN.  The constraint (\ref{wsnp}-2) ensures that the number of associated SNs to GN $ m $ is less than $ \chi_{m} $\footnote{$\chi_m $ depends on the underlying channel bandwidth and multiple access technique (e.g., FDMA, TDMA, and OFDMA) used by GNs to serve their associated SNs.}. Moreover, (\ref{wsnp}-3) guarantees that the gateways' average operational cost  is less than $\Gamma$ and (\ref{wsnp}-4) enforces that association with GN $ m $ is only possible when GN $ m $ is on. It is assumed that GNs have wired/wireless connection to an IoT controller which is responsible for solving the WSN-side problem and sending the result back to the GNs. 
\\ In addition to the WSN, a smart jammer node (JN) is also present. The JN operates reactively; it first overhears ongoing links,   then selects a set of sensors as the \textit{victims}, and finally attempts to disrupt the victims' links by sending the interfering signals. We assume the JN can detect the signal of SN $n$ by probability of $p_n$ and attempts to take its most malicious action by solving the following optimization problem:
\begin{align}
\begin{array}{lll}
\mathbf{P2:}& \min_{\mathbf{v}} \sum_{n=1}^{N}w_nv_{n}+\lambda\sum_{n=1}^{N}\rho_n v_n
\\&
\quad \text{subject to } v_n \in \{0,1\},\quad\quad n\in\mathcal{N}
\end{array}
\end{align} 
where $ \mathbf{v}=[v_{1},v_{1},\cdots,v_{N}] $, and $v_n$ is a binary variable which indicates whether the SN $ n$  $ $ is among the victims ($ v_n=1 $) or not ($ v_n=0 $). Moreover,
$w_n$ is the parameter \textit{learned} by the jammer to weigh the decision of selecting SN $n$ as the victim and defined as  
\begin{align}
w_n=&\tilde{Pr}(\text{SPKT($ n $) $| v_n $=1})-\tilde{Pr}(\text{SPKT($ n $) $| v_n $=0})
\end{align}
where $ \tilde{Pr}(.) $ stands for the \textit{jammer-measured} probability, and SPKT($ n $) is the event of successful packet transmission by the SN $n$. 
Therefore, the first term in the objective function of \textbf{P2} is to minimize the  WSN's average PDR from the \textit{jammer's perspective}. In the second term, $\rho_n=p_n\Omega$ is the average power that JN consumes to jam SN $ n $ where $\Omega$ is the power of the interfering signal. Accordingly, $\sum_{n=1}^{N}\rho_n v_n$ represents the average jamming power, and
$\lambda$ is a weight balancing the jammer's desire to consume less power against its intention to disturb the WSN as much as possible. The case in which  $\lambda=0$, represents a jammer without any power consumption concern. On the other hand, $\lambda\rightarrow\infty$ seizes the jammer operation due to  its power shortage.
\subsection{Stackelberg Game Formulation}
Although \textbf{P1} and \textbf{P2} look like separated problems, they are coupled with the coupling parameters of $p_{n,m}$ and $w_n$ in \textbf{P1} and \textbf{P2}, respectively. Specifically, for the $p_{n,m}$, we can write
\begin{align}
p_{n,m}=&v_nPr(\text{SPKT}(n,m)|v_n=1)\nonumber\\
&+(1-v_n)Pr(\text{SPKT}(n,m)|v_n=0)\nonumber\\
=&a_{n,m}v_n+b_{n,m}
\end{align} 
where $Pr(.)$ is the \textit{WSN-measured} (true) probability,  $ \text{SPKT}(n,m) $ denotes the event of successful delivery of a packet sent by SN $ n $ to GN $ m $, $a_{n,m}=(p_np^{(1)}_{n,m}-p_np^{(0)}_{n,m}-p^{(0)}_{n,m})\le0$, $ b_{n,m}=p^{(0)}_{n,m} $, $p^{(0)}_{n,m}=Pr(\text{SPKT}(n,m)|\text{SN $ n $ is not jammed})$,
and $p^{(1)}_{n,m}=Pr(\text{SPKT}(n,m)|\text{SN $ n $ is jammed})$.  On the other hand,
\begin{align}
\label{eq17abn1}
w_n=\sum_{m=1}^{M}(\tilde{p}^{(1)}_{n,m}-\tilde{p}^{(0)}_{n,m})x_{n,m},
\end{align}
where $\tilde{p}^{(0)}_{n,m}$ and $ \tilde{p}^{(1)}_{n,m} $ are the jammer-measured probability of SPKT$ (n,m) $ when $v_n=0$ and $v_n=1$, respectively. 
\\\indent \textit{Remark} 1: Equation (\ref{eq17abn1}) is general enough to model different scenarios based on the jammer's learning power. For instance, the jammer may simply assume that non-jammed transmissions are always successful and jammed ones are always failed (in this case $ w_n $ reduces to $ -\sum_{m=1}^{M}x_{n,m} $). On the other hand, a much more complex jammer may measure the WSN's signaling (e.g., ACK/NACK) and data transmissions for a long enough time to precisely learn $\tilde{p}^{(0)}_{n,m}=p_n{p}^{(0)}_{n,m}\hat{p}_m$ and $\tilde{p}^{(1)}_{n,m}=p_n{p}^{(1)}_{n,m}\hat{p}_m$ where $ \hat{p}_m $ is the probability that the jammer detects the ACK signals sent by GN $ m $.
\\\indent \textit{Remark} 2: The parameter $p_n$ (and $ \hat{p}_m $) models the locality of the JN. Some SNs maybe hidden to the JN ($p_n\approx0$) and some maybe in its complete detection scope ($p_n\approx1$). In general, $p_n$ depends on the state of the wireless channel between the JN and the SN $ n $, the sensitivity of JN's receiver, and the transmission power of the SN $ n $.
\\ 
The coupling among the \textbf{P1} and \textbf{P2} lays the foundation for casting the problem as a Stackelberg game. In game theory, the Stackelberg model is a strategic game in which two types of players pursue a hierarchy of actions: one player is the \textit{leader}, and the other is the \textit{follower}. The leader, as the name suggests, takes its action first. The follower observes the move and chooses the action that maximizes its payoff (i.e., the best response strategy). With a prior knowledge about the follower's reaction, the leader can opt its own optimum strategy. This leader's optimal strategy, in combination with the corresponding follower's best response, is a Stackelberg (Nash) Equilibrium (SE) of the game. In our model, the jammer selects its victims based on the sensors'  contributions in WSN's average PDR (estimated by the jammer).
Therefore, we can model the WSN's association and gateway scheduling problem in the presence of a smart jammer as a Stackelberg game, $\mathcal{G^A}$. 

\begin{definition}
The Stackelberg Association game $\mathcal{G^A}$ is defined as a triplet $(P,S,U)$. $P$ is the set of players including WSN and jammer, where WSN is the leader and jammer is the follower. $S$ is the set of strategies, where the strategy of the leader is the association and gateway scheduling scheme and the strategy of the follower is the set of victims. Finally $U$ is the set of payoff functions, where \textbf{P1} and \textbf{P2} represent the utility functions of the leader and the follower, respectively.
\end{definition}

In the following section we analyze game $\mathcal{G^A}$ and obtain Nash equilibria of this game. 


\section{Optimal Association And Gateway Scheduling}
\label{sec_GameAnalysis}

In order to analyze game $\mathcal{G^A}$, for a given strategy of the WSN, we derive the best response strategy of
the jammer. Then, based on the knowledge of jammer's strategy, we formulate the optimal strategy of the WSN which identifies a Stackelberg equilibrium (SE) of the game.
For a given WSN's strategy $\mathbf{x}=[x_{1,1},x_{1,2},\cdots,x_{N,M}]$, the jammer's best response is obtained by solving \textbf{P2} while treating $ \mathbf{x} $ as known parameters. The following lemma presents how we can compute the jammer's best response strategy.
\begin{lemma}\label{lemma_BRJammer}
In game $\mathcal{G^A}$, let $ \mathbf{x} $ denote a given strategy of the WSN. Then the
best response of the jammer is
\begin{align}
\label{18abn1}
v_n = \begin{cases}
\begin{array}{l l}
0& w_n+\lambda\rho_n>0\\
1& w_n+\lambda\rho_n\leq0
\end{array}
\end{cases}
\end{align}
\end{lemma}
\begin{IEEEproof}
Plugging the given $ \mathbf{x} $ into \textbf{P1} makes all $ w_n $, $ n\in\mathcal{N} $, constant and the objective function becomes a linear function of $ \mathbf{v} $. As a result, the optimum choice for the binary variable $ v_n $ is 0 when its coefficient, $ w_n+\lambda\rho_n $, is positive and 1, otherwise.
\end{IEEEproof}

Assuming that the WSN knows the jammer's best response strategy, given in (\ref{18abn1}), it can compute its
optimal strategy, by solving the following optimization problem.
\begin{align}
\begin{array}{llr}
\label{wsnp3}
\mathbf{SE:}& \max_{\mathbf{x,v}} \sum_{m=1}^{M}\sum_{n=1}^{N}(a_{n,m}v_n+b_{n,m})x_{n,m}&\vspace{1mm}\\\vspace{1mm}
&\text{subject to}\\&\quad
\sum_{m=1}^{M}x_{n,m}\leq1,\quad\hspace{0.5mm} n\in\mathcal{N}&\text{(\ref{wsnp3}-1)}\\
&\quad
\sum_{n=1}^{N}x_{n,m}\leq \chi_{m},\quad\hspace{-1.5mm} m\in\mathcal{M}&\text{(\ref{wsnp3}-2)}\\
&\quad
\sum_{m=1}^{M} c_my_m
\leq \Gamma&\text{(\ref{wsnp3}-3)}\\
&\quad
x_{n,m}\leq y_m,\quad m\in\mathcal{M}, ~n\in\mathcal{N}&\text{(\ref{wsnp3}-4)}\\
&\quad v_n=\Pi\left(-w_n-\lambda\rho_n\right),\quad n\in\mathcal{N}&\text{(\ref{wsnp3}-5)}\\
&\quad
y_m,x_{n,m},v_n\in\{0,1\},\quad m\in\mathcal{M}, ~n\in\mathcal{N}&\text{(\ref{wsnp3}-6)}
\end{array}
\end{align}
where $\Pi(.)$ is the unit step function. The problem \textbf{SE} is an integer non-linear optimization problem. But in view of the following lemma,  an integer linear programming (ILP) also exists which simplifies finding the WSN's optimal solution using commercial ILP solvers.

\begin{lemma}\label{lemma_BRWSN} 
The optimal strategy of WSN is the solution of the \textbf{ILP-SE} problem.
\begin{align}
\begin{array}{llr}
&\mathbf{ILP-SE:}&\\
& \max_{\mathbf{x,z}} \sum_{m=1}^{M}\sum_{n=1}^{N}a_{n,m}z_{n,m}+b_{n,m}x_{n,m}&\vspace{1mm}\\\vspace{1mm}
&\text{subject to}\\&\quad
\sum_{m=1}^{M}x_{n,m}\leq1,\quad\hspace{0.5mm} n\in\mathcal{N}&\\
&\quad
\sum_{n=1}^{N}x_{n,m}\leq \chi_{m},\quad\hspace{-1.5mm} m\in\mathcal{M}&\\
&\quad
\sum_{m=1}^{M} c_my_m
\leq \Gamma&\\
&\quad
x_{n,m}\leq y_m,\quad m\in\mathcal{M}, ~n\in\mathcal{N}&\\
&\quad v_n\ge-\lambda\rho_n-\sum_{m=1}^{M}(\tilde{p}^{(1)}_{n,m}-\tilde{p}^{(0)}_{n,m})x_{n,m},\quad n\in\mathcal{N}&\\
&\quad
z_{n,m}\le (x_{n,m}+v_n)/2,\quad m\in\mathcal{M}, ~n\in\mathcal{N}&
\\
&\quad
z_{n,m}\ge v_n+x_{n,m}-1,\quad m\in\mathcal{M}, ~n\in\mathcal{N}&
\\
&\quad
y_m,x_{n,m},v_n,z_{n,m}\in\{0,1\},\quad m\in\mathcal{M}, ~n\in\mathcal{N}&\\
\end{array}
\end{align}
where $ \mathbf{z}=[z_{1,1},z_{1,2},\cdots,z_{N,M}] $.
\end{lemma}

\begin{IEEEproof}
We can derive \textbf{ILP-SE} from \textbf{SE} taking the following two steps.\\
Step 1: We replace the non-linear condition (\ref{wsnp3}-5) in \textbf{SE} by 
\begin{align}
\label{linc}
v_n\ge-\lambda\rho_n-\sum_{m=1}^{M}(\tilde{p}^{(1)}_{n,m}-\tilde{p}^{(0)}_{n,m})x_{n,m}
\end{align}
The reason is that for the case in which RHS of (\ref{linc}) is positive, both (\ref{linc}) and (\ref{wsnp3}-5) yield $v_n=1$ noting that $ v_n\in\{0,1\} $. On the other hand, for the case in which RHS of (\ref{linc}) is negative, (\ref{wsnp3}-5) results in $ v_n=0 $, but  (\ref{linc}) lets $v_n$ be free ($v_n$ can be either 0 or 1). However, note that the objective function in \textbf{SE} favors $v_n=0$  as $ a_{n,m}\le0 $.  So again in this case replacing  (\ref{wsnp3}-5) by (\ref{linc}) makes no difference.
\\Step 2: The nonlinear terms in the objective function of \textbf{SE}, in the form of $v_nx_{n,m}$, can be replaced by $z_{n,m}$ with the following linear conditions:
\begin{align}
\label{18abn2}
&z_{n,m}\le (x_{n,m}+v_n)/2\\
\label{18abn3}
&z_{n,m}\ge v_n+x_{n,m}-1\\
\label{18abn4}
&z_{n,m}\in\{0,1\}
\end{align} 
This replacement can also be easily verified by observing that conditions (\ref{18abn2}), (\ref{18abn3}), and (\ref{18abn4}) enforce $ z_{n,m}=v_nx_{n,m} $ in all cases where  $ (v_n,x_{n,m})=(0,0) $, $ (v_n,x_{n,m})=(0,1) $, $ (v_n,x_{n,m})=(1,0) $, and $ (v_n,x_{n,m})=(1,1) $.
\end{IEEEproof}
Using Lemma~\ref{lemma_BRJammer} and \ref{lemma_BRWSN}, we can now compute the Nash equilibrium of game $\mathcal{G^A}$ by the following theorem. 

\begin{theorem}
The strategy pair $ \mathbf{(x^*,v^*)} $ is the Stackelberg Equilibrium of the game $\mathcal{G^A}$, where
\begin{align}
\mathbf{(x^*,z^*)}= \arg\{\text{\textbf{ILP-SE}}\}
\end{align}
and $ v^*_nx^*_{n,m}=z^*_{n,m} $, $ n\in\mathcal{N} $, and $ m\in\mathcal{M} $.
\end{theorem}

Note that the jammer needs to know  $ w_n, n\in\mathcal{N} $  to play its best response, $ \mathbf{v^*} $. In practice, the smart jammer may perform learning, measurements and/or eavesdropping the signaling channels in order to obtain  such information. On the other hand, to calculate $\mathbf x^*$, the WSN also may conduct machine learning methods to detect and identify the strategy of the jammer \cite{7158697, 5683729}. Moreover, \textbf{ILP-SE} provides an achievable upper bound on  the WSN's performance considering different association and gateway scheduling schemes. Therefore, \textbf{ILP-SE} may serve as an analytical tool for analyzing and designing multi-gateway WSNs in jamming environments. 

\section{Simulations}
\label{sec_Simulation}

We now present simulation results that illustrate the performance of multi-gateway WSNs against smart jammers.
\begin{figure}[t]
\centering
	\includegraphics[scale=0.5]{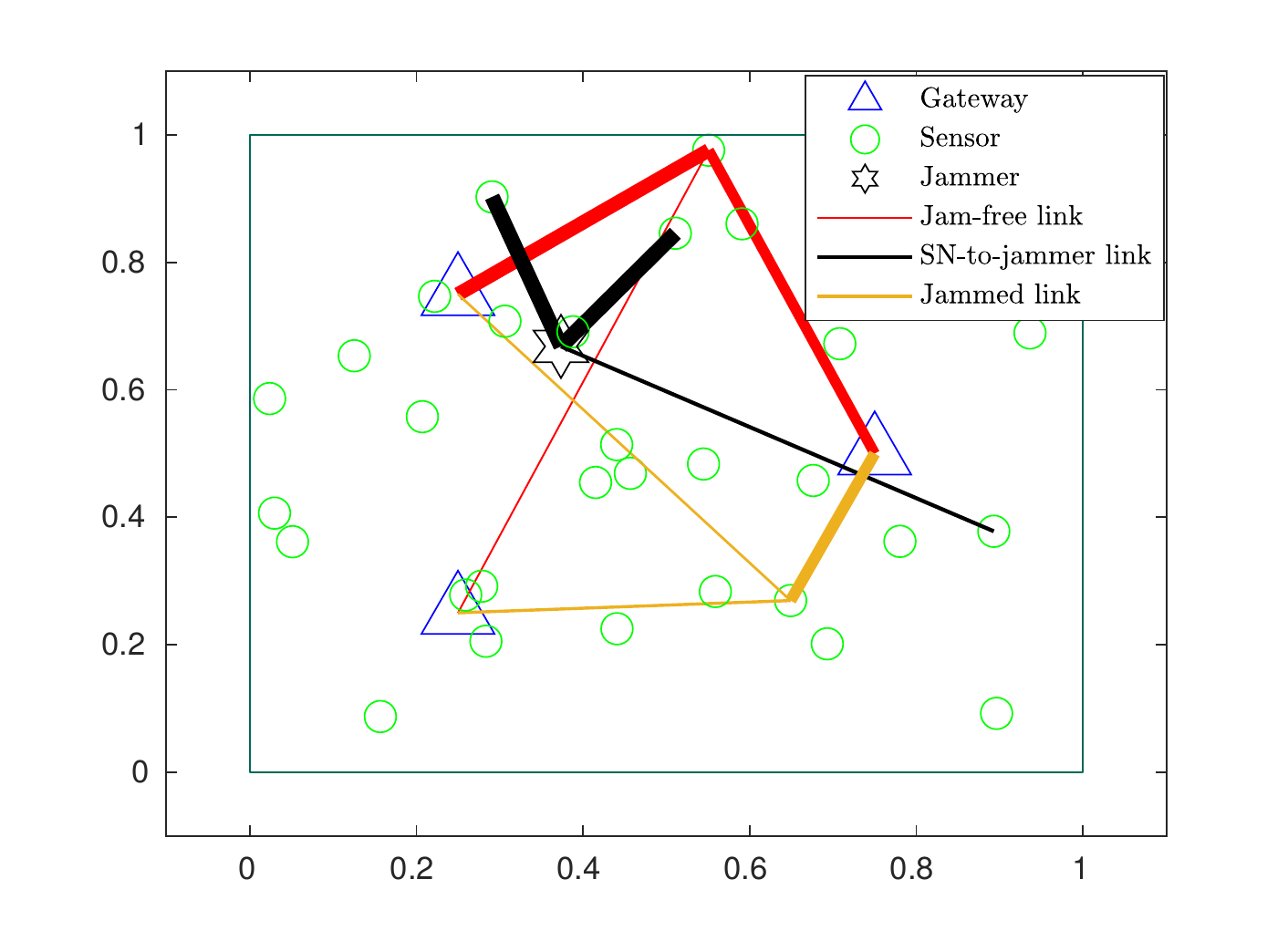}
	\caption{An instance of a system model generated for simulation ($M=3, N=30$). The system can be abstracted as a graph with three types of edges (links): (a) the red links between SNs and GNs \textit{without jamming} (the width of the edge between SN $ n $ and GN $ m $ is proportional to $ p^{(0)}_{n,m} $), (b) the black links between SNs and the jammer (the width of the edge between SN $ n $ and the jammer is proportional to $ p_{n} $), and (c) the orange links between SNs and GNs \textit{under jamming} (the width of the edge between SN $ n $ and GN $ m $ is proportional to $ p^{(1)}_{n,m} $). For the sake of clarity just a few number of links are depicted.}
	\label{20abnfif1}
\end{figure}
We consider a Rayleigh flat fading channel model for wireless links. For the successful packet transmission over such a  link (with no jamming), we can write
\begin{align}
\label{e1_21far}
Pr(success) = 1-p_{out}(R) = \exp(\frac{-(2^R-1)}{\text{SNR}}) 
\end{align}
where SNR is the signal to noise ratio, $ R $ represents the bit rate, $p_{out}(R)=Pr\{R>\log(1+|h|^2 \text{SNR})\}$ is the outage probability, and $h\sim CN(0,1)$ is a zero-mean unit-variance complex Gaussian variable which models the Rayleigh fading \cite{goldsmith2005wireless}.
On the other hand, in wireless links, the  SNR can be modeled as being proportional to $ d^{-\alpha} $ (i.e., SNR$\propto d^{-\alpha}$), where $d$ is the distance between the transmitter and the receiver, and $\alpha$ is the path-loss exponent \cite{goldsmith2005wireless}. 
 Assume that  $ Pr(success) = p^{(0)}_0$ over a link with the length of  $d_0$. As a result, based on (\ref{e1_21far}), for the link between the SN $ n $ and GN $ m $, we have
\begin{align}
\label{e3_21far}
p^{(0)}_{n,m} = \left({p_{0}^{(0)}}\right)^{(\frac{d_{n,m}}{d_0})^\alpha}
\end{align}
where $ d_{n,m} $ is the distance between SN $ n $ and GN $ m $. Similarly, we can compute $p^{(1)}_{n,m}$ and $p_n$ by,
\begin{align}
\label{19abne1}
&p^{(1)}_{n,m} = \left({p_{0}^{(1)}}\right)^{(\frac{r_{n,m}}{r_{0}})^\alpha},\quad
p_n={p_0}^{(\frac{D_{n}}{D_{0}})^\alpha}
\end{align}
where $p_{0}^{(1)}$ is the probability of successful packet transmission from a reference point when its signal is jammed and its distance to the destination over that to the jammer is $r_0$. Moreover, $r_{n,m}$ is $d_{n,m}$ divided by the distance between SN $ n $ and the jammer. In Equation (\ref{19abne1}), $p_0$ is the probability that the jammer detects the transmission of a sensor located at its $D_{0}$ distance, and $ D_{n} $ denotes the distance between the SN $ n $ and the jammer, and we have assumed that jammer can detect the transmission of SN $n$, if its SNR at the jammer location is large enough.  
In our simulation, the jammer and the SNs are randomly dispersed within the unit square. Unless otherwise specified, we also assume $\alpha=4$, $p_0^{(0)}=0.7$, $d_0=0.5$, $p_0^{(1)}=0.1$,  $r_0=1$, $p_0=0.5$,  $D_0=0.5$, $\Omega = 1$, $\Gamma = 2$, $c_m = 1$ and $\chi_m=10, m\in\mathcal{M}$. Fig. \ref{20abnfif1} depicts an instance of the randomly generated system model with 3 GNs and 30 SNs. 

\begin{figure}
\centering
	\includegraphics[scale=0.5]{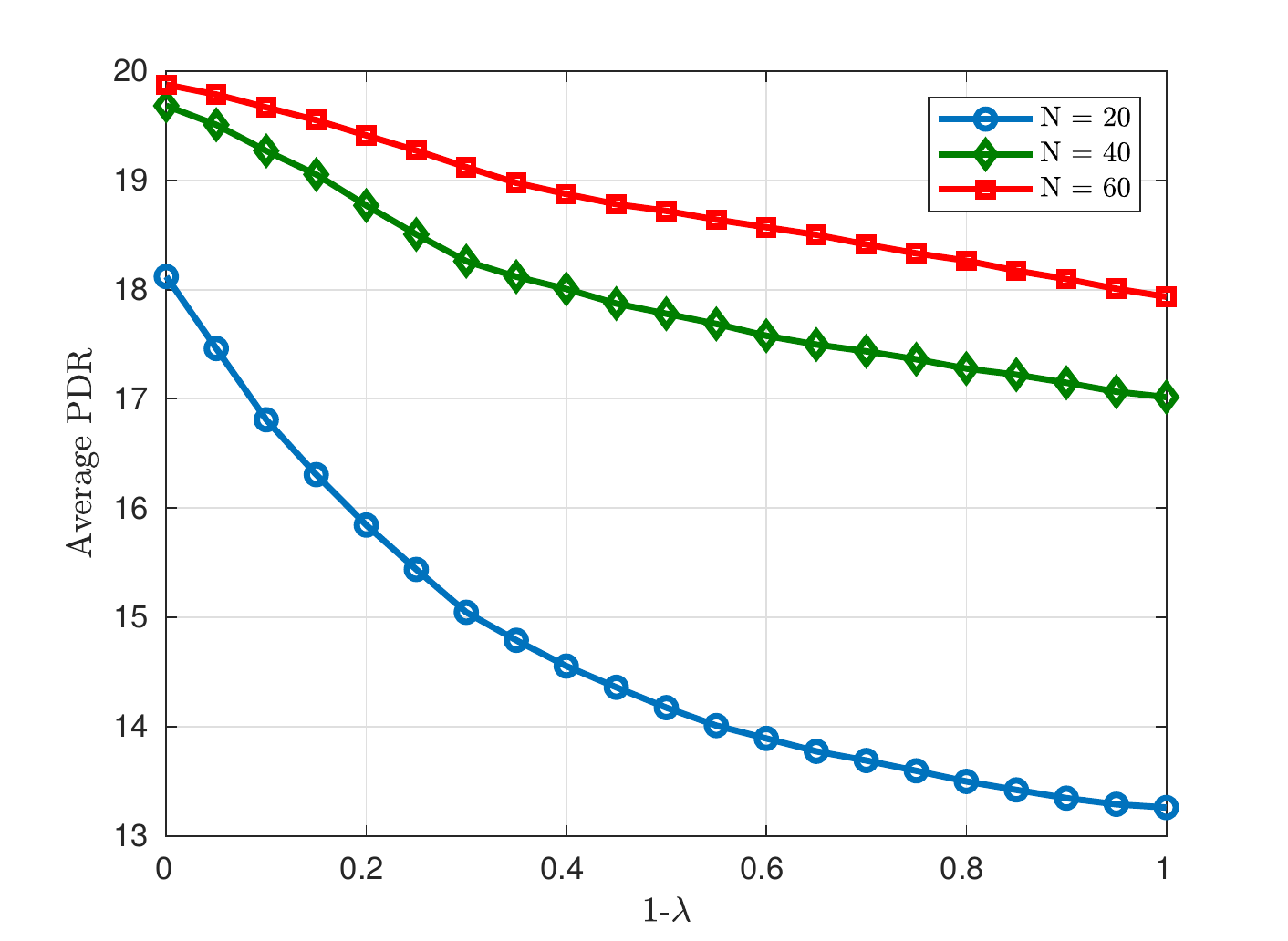}
	\caption{The performance of WSN with two gateways and different number of SNs.}
	\label{21abnf1}
\end{figure}
In our simulations, we have used Matlab's mixed-integer linear programming (intlinprog) solver to find the solution for the ILP problems. Fig. \ref{21abnf1} shows the  performance, in terms of average packet delivery rate (PDR), of the WSN with 2 GNs ($ M=2 $) and different number of SNs with respect to the jammer's power (the horizontal axis indicates $1- \lambda$ ). Each data point represents the average performance achieved at the Stackelberg equilibrium (i.e., the solution of \textbf{ILP-SE} problem). The results are reported in  the form of the average over 100 experiments (iterations). It is observed that PDR decreases  as the jammer's power increases. Specifically, in the low power region of the jammer, PDR approaches to its maximum value of 20 as the maximum number of associated SNs to each GN is 10 ($ \chi_1=\chi_2=10 $). Moreover, it is observed that as the number of SNs increases the effect of the jammer on the WSN's performance decreases. In particular, the performance of the WSN changes by 27\%, 14\%, and 10\% when the number of SNs are 20, 40, and 60, respectively. The reason is that as the number of SNs grows the WSN has more options and can select those which their transmitted signals are less polluted by the JN.
\begin{figure}[t]
\centering
	\includegraphics[scale=0.5]{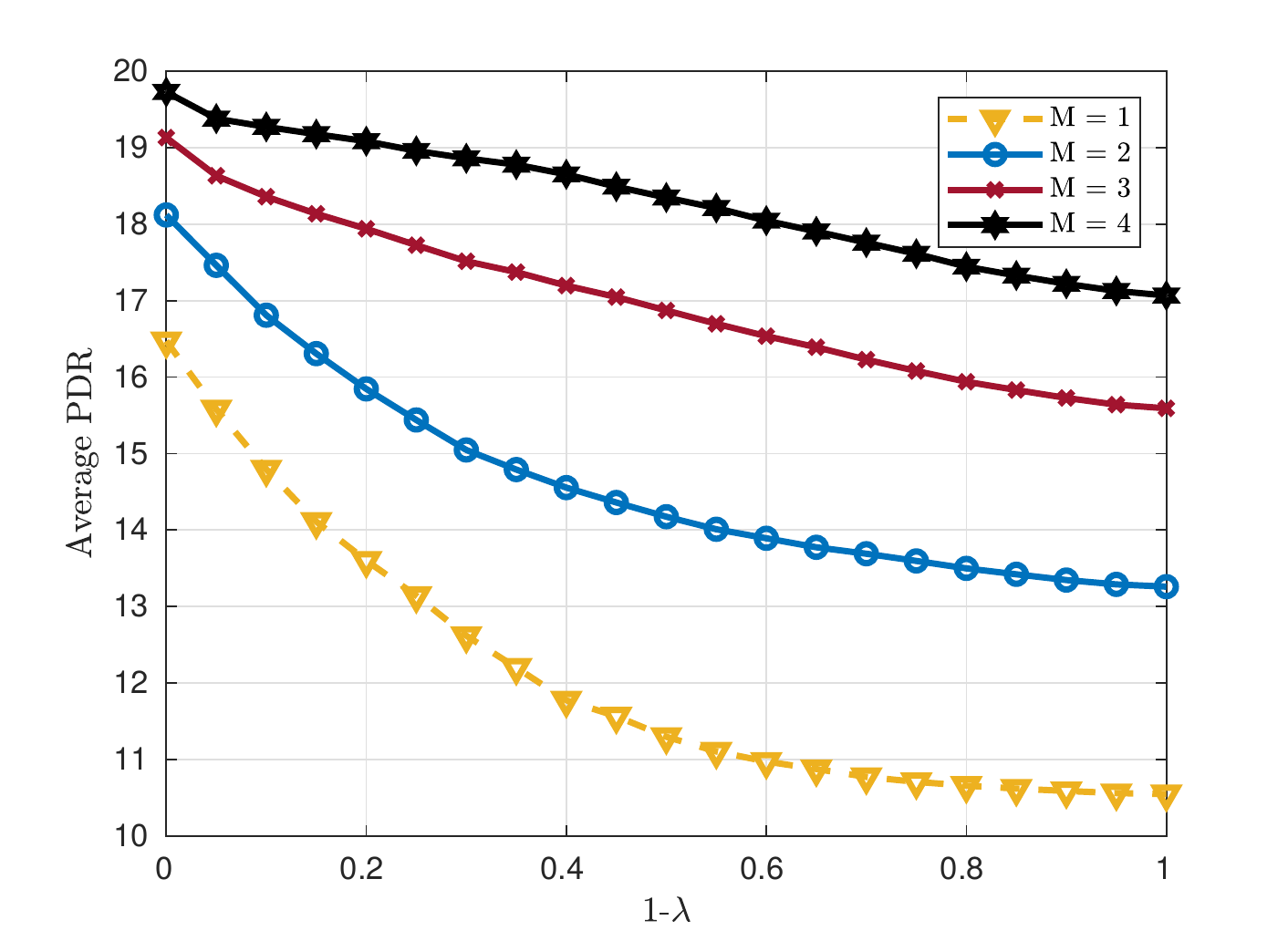}
	\caption{The performance of WSNs with different number of gateways.}
	\label{21abnf2}
\end{figure}

\begin{figure}[t]
\centering
	\includegraphics[scale=0.50]{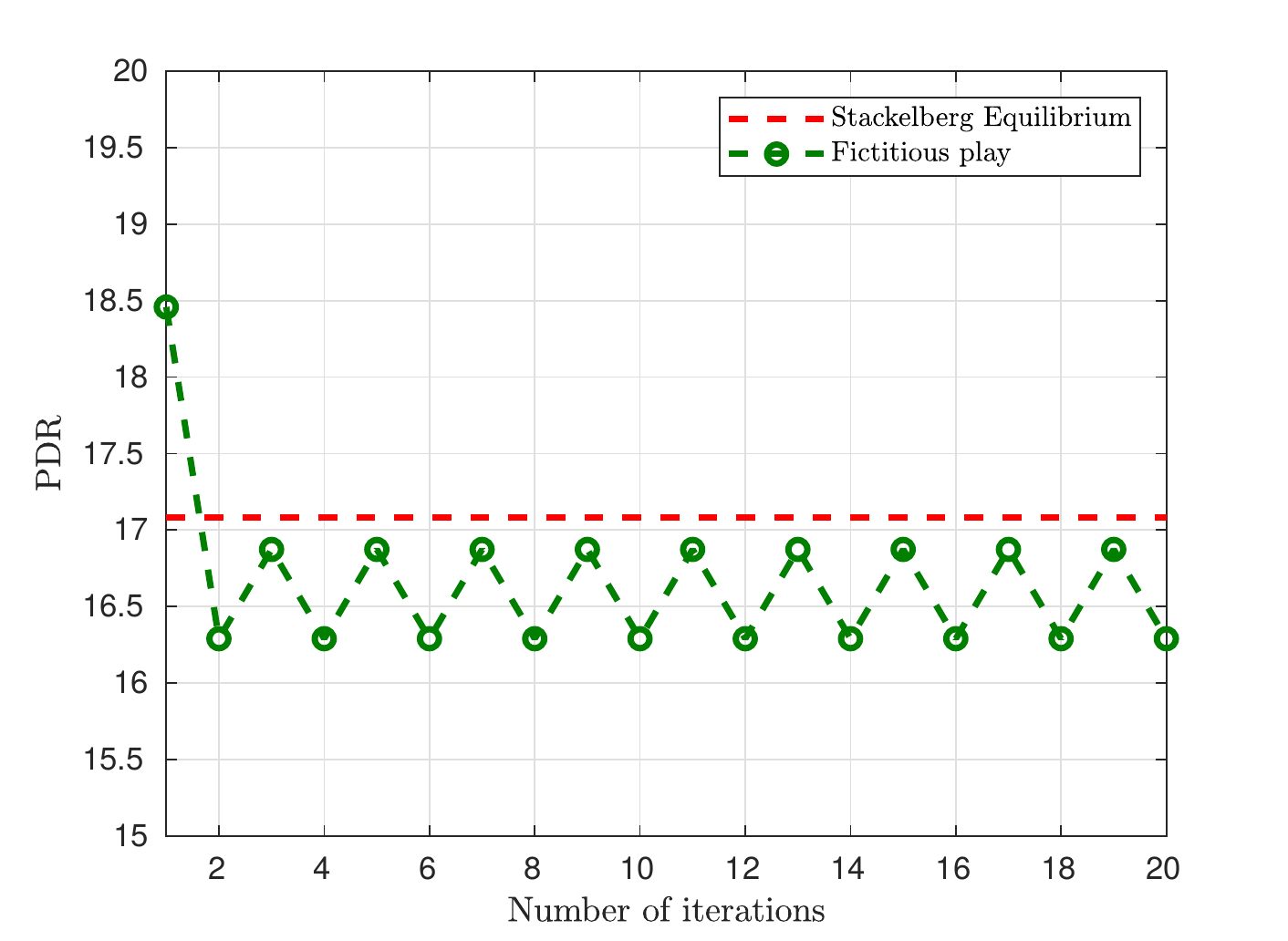}
	\caption{A sample of the performance of the system in which the WSN and jammer take the fictitious play strategy}
	\label{21abnf3}
\end{figure}

Fig. \ref{21abnf2} compares the performance of the WSNs with different number of gateways for the same operation cost that allows switching on two gateways ($\Gamma=2, c_m=1, m\in\mathcal{M}$). As the benchmark, the performance of a single gateway system ($M =1$) at the center which can serve up to 20 associated SNs ($\chi_1 = 20$) is also included. The number of SNs in all cases is 20. For the two gateway scenario the GNs are positioned at coordinate (0.25,0.5) and (0.75,0.5). On the other hand, for the three gateway scenario, GNs are located at (0.25,0.25), (0.25,0.75), and (0.75,0.5), and for the four gateway scenarios, GNs ar in the center of each quadrant. As the figure shows the performance enhances (and more robust to jamming power level) as the number of GNs are increased, because with more GNs, the WSN can circumvent the jammer by scheduling those GNs to be on, which jammer is less able to disturb their received signals. Moreover, it is observed that the performance of the two-gateway scenario is better up to 26\% with respect to the single-gateway scenario. This performance gap climbs up to 60\% considering the sccenario with four gateways. On the other hand, comparing Fig. \ref{21abnf1} and Fig. \ref{21abnf2} suggests that by using 4 GNs, instead of 2, we can reduce the number of sensors from 60 down to 20 and still obtain almost the same performance.  

Finally, Fig. \ref{21abnf3} illustrates the performance of the system when the WSN and the jammer, in turn, play their best response in reaction to the other (also known as the  \textit{fictitious play}  strategy). The number of GNs is 2, the number of SNs is 20, $\lambda=0.75$, and the PDR of the WSN in the first 20 interactions with the jammer is depicted. As it is evident, the best response strategy may not converge (and in the case of convergence, it will converge to the Stackelberg equilibrium). 

\balance

\section {Conclusion}

Low-power sensor networks are all vulnerable to jamming attacks. In this paper, we have studied the interaction between jammer and wireless sensor networks with several gateways. Using a game-theoretic model, we have derived the optimal strategies of attackers and sensor networks in a Stackelberg setting. Our simulation results show that how increasing the number of gateways can mitigate the jamming attack. In other words, we have shown how we can decrease the impact of jamming attacks by deploying more gateways. A particularly interesting result is that with only three extra gateways, we can increase the network's throughput by 60\% under same jamming attack. The proposed model is the first step towards a deeper understanding of learning in such security scenarios where the attackers and defenders have sequential interactions. For future work, we intend to evaluate such attacks and countermeasures with various learning approaches.

\bibliography{wsnref} 

\bibliographystyle{ieeetr}	

\end{document}